\documentclass[12pt]{article}

\usepackage{scicite}
\usepackage{graphicx} 

\usepackage{times}

\newenvironment{sciabstract}{%
\begin{quote} \bf}
{\end{quote}}

\newcounter{lastnote}
\newenvironment{scilastnote}{%
\setcounter{lastnote}{\value{enumiv}}%
\addtocounter{lastnote}{+1}%
\begin{list}%
{\arabic{lastnote}.}
{\setlength{\leftmargin}{.22in}}
{\setlength{\labelsep}{.5em}}}
{\end{list}}

\title{Cosmological Magnetic Field: a fossil of density perturbations
in the early universe} 

\author
{Kiyotomo Ichiki,$^{1,2,3\ast}$ Keitaro Takahashi,$^{3,4}$ Hiroshi
Ohno,$^{5}$ \\ Hidekazu Hanayama,$^{1,6}$ Naoshi Sugiyama$^{1}$\\
\\
\normalsize{$^{1}$National Astronomical Observatory of Japan, Mitaka, Tokyo 181-8588, Japan}\\
\normalsize{$^{2}$Kavli Institute for Cosmological Physics, University
of Chicago, Chicago, IL 60637, USA}\\
\normalsize{$^{3}$Research Fellows of Japan Society for the Promotion of Science}\\
\normalsize{$^{4}$Department of Physics, Princeton University,
Princeton, NJ 08544, USA}\\
\normalsize{$^{5}$Corporate Research and Development Center, Toshiba
Corporation, Kawasaki 212-8582, Japan}\\
\normalsize{$^{6}$Department of Astronomy, University of Tokyo, 7-3-1 Hongo,
 Bunkyo-ku, Tokyo 113-0033, Japan}\\
\\
\normalsize{$^\ast$To whom correspondence should be addressed; E-mail:  ichiki@th.nao.ac.jp.}
}

\date{Published in {\it Science}, {\bf 311}, 827-829, 2006}

\begin{document} 


\baselineskip24pt


\maketitle 

\begin{sciabstract}
 The origin of the substantial magnetic fields that are found in
 galaxies and on even 
 larger scales, such as in clusters of galaxies, is yet unclear.
 If the second-order couplings between photons and electrons are
 considered, then 
 cosmological density fluctuations, which explain the large
 scale structure of the universe, can also produce magnetic fields on
 cosmological scales before the epoch of recombination.
 By evaluating the power spectrum of these cosmological magnetic fields
 on a range of scales, we show here that magnetic fields of $10^{-18.1}$
 gauss 
 are generated at a $1$ megaparsec  scale and can be even stronger at
 smaller 
 scales ($10^{-14.1}$ gauss at $10$ kiloparsecpc).
 These fields are large enough to seed magnetic fields in galaxies and
 may therefore have affected primordial star formation in the early
 universe.
\end{sciabstract}


Conventional models for the generation of large-scale magnetic
fields are mostly classified into two categories: astrophysical
and cosmological mechanisms.
Astrophysical mechanisms---often involving the Biermann battery effect in
which magnetic fields are generated from an electric current driven by
the rotation of the system \cite{Biermann50}---can explain the small-scale
amplification of fields, such as in stars or in supernova explosions
\cite{1962ApJ...136..615M,Hanayama05}.
However, these mechanisms do not fully explain fields on larger cosmological
scales, such as those known to exist in galaxies and clusters of
galaxies.
Reconnecting the magnetic field lines might increase the coherence
length from the size of stars to that of galaxies (tens of kpc) although
there is still no convincing evidence to favor 
this scenario (reviewed in \cite{Widrow02}).
Or perhaps large-scale magnetic fields were directly generated in
the early universe, several 100 million years after the Big Bang, when
the universe was reionized \cite{Gnedin00} or protogalaxies were formed
\cite{Kulsrud97,Davies00}. However the models still remain uncertain
because of the lack of observations of a high redshift universe. 

On the other hand, cosmological mechanisms based on inflation
\cite{Guth:1980zm,Sato:1980yn} have no difficulty in accounting for the
length of coherence; the accelerating expansion of the 
universe stretches small-scale quantum fluctuations to scales that
can exceed the causal horizon. 
However, because standard electromagnetic fields are conformally coupled
to gravity, magnetic fields simply dilute away as the universe expands.
Eventually the amplitudes of the magnetic fields
become negligibly small at the end of inflation. To produce
substantial primordial magnetic fields during inflation, new coupling---
such as exotic coupling of electromagnetic fields to non-standard
particles \cite{Ratra:1991bn,Lemoine:1995vj,Bamba:2003av} or gravity
\cite{Turner:1987bw}--- must be introduced. This new coupling must
amplify magnetic fields against cosmological expansion. 
Therefore, the nature of generated
magnetic fields, such as their amplitude or spectrum, depends strongly
on the assumptions built into standard cosmology or particle
physics. 
Moreover, it is argued that almost all the models that generate
magnetic fields at the inflationary epoch are ruled out, because they would
produce a large amount of gravitational waves before the Big Bang
nucleosynthesis, which then make cosmic expansion faster to bring an
overproduction of helium nuclei in the universe \cite{Caprini:2001nb}.

In addition to the astrophysical and cosmological mechanisms,
there is a third category for the generation of large-scale magnetic
fields.
Small density fluctuations during the cosmological recombination of
hydrogen atoms inevitably induce magnetic fields.
Compton and Coulomb scatterings are so efficient that photons,
protons and electrons are approximated to a tightly coupled fluid. 
If these three kinds of fluids moved in exactly the same way, magnetic
fields could not be generated. However, because photons scatter off
electrons preferentially compared with protons, small differences in
velocity between protons and electrons are generated, which yields an
electric current \cite{Harrison70,Lesch95}. Moreover, we show 
that the anisotropic pressure of photons pushes the electrons in a
different way from the protons (eq. S2). The rotation of the electric
current thus generates magnetic fields.
However, the rotation (or vector) mode of perturbations in the linear
order is known to be damped away in the expanding universe.
Therefore it is essential to consider the second-order couplings in the
Compton scattering term
\cite{Berezhiani04,Matarrese:2004kq,Takahashi:2005nd}. 
The magnetic fields generated through this process are
correlated with temperature fluctuations at the recombination epoch
because the electric current is associated with the density perturbations of
photons (Fig. 1).

There are three main contributions to the generation of magnetic fields
(eq. S2): (i) the baryon-photon slip term, (ii) the vorticity difference
term, and (iii) the anisotropic pressure term.
These terms are derived from the fact that electrons are pushed by photons
through Compton scattering when velocity differences exist between
them or when there is anisotropic pressure from photons.
We derive here the power spectrum of magnetic fields (eq. S19), 
and then perform a numerical calculation to evaluate it.
The power spectrum of magnetic fields $S(k)$ is defined by the expected
variance of the Fourier component of magnetic fields ${\vec B}(\vec{k})$ as
$S(k) \equiv \left<\left|{\vec B}(\vec{k})\right|^2 \right>$, where
$\vec{k}$ is the wave vector. The component of the field with
characteristic wavelength scale $\lambda$ can then be derived through
$B_\lambda \approx 
\sqrt{[k^3 S(k)/(2\pi^2)]}$ with $\lambda = 2\pi/k$.
We consider a standard cosmological model \cite{Bennett:2003bz} which
consists of photons, baryons, cold dark matter, neutrinos, and
the cosmological constant, and we fix all the cosmological parameters to the
standard values (eq. S26). 
The density perturbations of these parameters were solved numerically for
a range of scales from $10$ kpc up to $10$ Gpc, and they were then
integrated to obtain $S(k)$ (eq. S19).
We found that the field strength of generated magnetic fields at
the time of cosmological recombination can be as large as $10^{-18.1}$ G
at 1Mpc comoving scale, and it becomes even larger at smaller scales 
($10^{-14.1}$ G at 10 kpc) (Fig. 2). 
After cosmological recombination, no magnetic fields would be generated,
because most of the electrons were combined into hydrogen atoms and Compton
scattering was no longer efficient.
This means that the fields presently have an amplitude of $10^{-24.1}$ G
at $1$Mpc ($10^{-20.1}$ G at $10$kpc) because magnetic fields decay
adiabatically as the universe expands after their generation. 
The field strength is large enough to seed the galactic magnetic fields
required by the 
dynamo mechanism, which is typically of the order of $10^{-20}$ to
$10^{-30}$ G at around the $10$ kpc scale \cite{Widrow02,Davis:1999bt}.

Over the range of scales calculated, 
the generated magnetic fields  increase monotonically with decreasing
scale.
We found that the field has a spectrum $S(k)\propto k^{4}$ at scales
larger than $\sim 10^{2.5}$ Mpc, which corresponds to super-horizon scales at 
recombination; $S(k)\propto k^{0}$ at intermediate scales ($10^{2.5}$
Mpc $<\lambda <10^{1.5}$ Mpc); and  $S(k)\propto k^{1}$ at
scales smaller than $\sim 10$ Mpc, where the contribution from the
anisotropic stress of photons 
dominates. This means that the field strength $B$ is proportional to
$k^2$ at scales smaller than $1$Mpc.
If the primordial power spectrum of density fluctuations is given
by a simple power law as predicted by inflation \cite{Starobinsky:1982ee},
our result implies that magnetic  
fields with strength $B \approx 10^{-12.8}$ G arise on a
$100$ pc comoving scale  at $z \approx 10$ (where $z$ is the
cosmological redshift). 
This value helps us to understand the evolution of structures in the
high-redshift universe, because those magnetic fields would be strong
enough to trigger a magneto-rotational instability in the accretion
disks surrounding very first stars (population III stars), and it affect
the transport of their angular momentum \cite{Maki:2004zt}.
The transport of angular momentum plays an important role for the
accretion of matter onto protostars. A typical mass scale of 
population 
III stars is key \cite{Maki:2004zt} for the early reionization and
chemical evolution of the universe; therefore,  cosmologically generated
magnetic 
fields should be one of the essential ingredients in the model of
structure formation in the high-redshift universe.

The behavior of the power spectrum $S(k)$ can be understood
by considering the spectra of source terms (eq. S3) at each redshift
from the deep radiation-dominated era to cosmological recombination
(Fig. 3).  We found that at recombination, both baryon-photon slip and
anisotropic stress contribute almost at the same order of magnitude
around horizon scales.  
The contributions from the earlier epochs are dominated by anisotropic
stress of photons, and they give rise to a larger amount of magnetic
fields at smaller scales. 
Because velocity differences between electrons and
photons are suppressed when energy density of radiation dominated
in the early universe, anisotropic stress of photons could not be
negligible at small scales. We also found that magnetic fields are mainly 
generated when the baryon-photon fluids undergo acoustic oscillation
after crossing the horizon [Fig. 3, left panel, (2)] until
photon diffusion processes \cite{Silk:1967kq}  erase the perturbations
[Fig. 3, left, (3) and (4)]. 
The generated spectrum of magnetic fields was obtained by the 
non-linear convolution and time integration of these spectra of source
terms (eq. S19), but approximately given by the superposition of them at
each redshift.

Because the creation of magnetic fields mainly occurs when the modes of density
perturbations with the corresponding scale enter the cosmic horizon
and become causally connected (Fig. 3), the magnetic fields should
exist at small scales below $\sim 10$ Mpc even where the Silk damping
effect by diffusion of photons has swept away the density perturbations
at the last scattering epoch. Thus, in 
principle, the detection of magnetic fields below the $\sim 10$ Mpc
scales calculated here would tell us about density
perturbations in photons (and baryons) in the early universe even at
scales smaller than 
the diffusion scale at recombination. In this sense, the magnetic field
generated by this mechanism can be regarded as a fossil of density
perturbations in the early universe, whose signature in photons and
baryons has been lost. Therefore, this result provides the 
possibility of probing observations on how density perturbations in
photons have evolved and been swept away at these small scales where no
one can, in principle, probe directly through photons.

The amplitude of the cosmologically generated magnetic fields is
too small to be observed directly through polarization effects or
synchrotron emission. However,  magnetic fields with
such small amplitude may be detected by gamma-ray burst observations
\cite{1995Natur.374..430P} through 
the delay of the arrival time of gamma-ray photons due to
the magnetic deflection of high energy electrons responsible
for such gamma-ray photons
\cite{Waxman:1996zc,Razzaque:2004cx}. 
We suggest that, because the weak magnetic fields 
should inevitably be generated from cosmological perturbations as
presented here, and because they should exist all over the universe even
in the intercluster 
fields, then the weak fields should be detectable by future high-energy 
gamma-ray experiments, such as GLAST (the Gamma Ray Large Area Space
Telescope) \cite{Gehrels:1999ri}. 

Although the power of $B$ increases as $k^2$ on small scales (Fig. 2),
the diffusion due to Coulomb scattering between electrons and protons
damps the magnetic fields around $k\sim 10^{12}$
Mpc$^{-1}\left[\frac{1+z}{10^4}\right]^{7/4}$. 
Therefore, the energy
 density in magnetic fields remains finite.  
Because magnetic fields at
smaller scales result from density perturbations in the early
universe, one needs to take into consideration
the high-energy effects neglected in the collision term: e.g.,
relativistic corrections for the energy of electrons and
weak interactions along with the Compton scattering.
These effects would become important when the temperature of the
universe was above $\sim 1$ MeV, which corresponds to the comoving wave
number $\sim 10^5$ Mpc$^{-1}$.
These effects can be safely neglected at the scales considered here.


\bibliography{mag}
\bibliographystyle{Science}


\begin{scilastnote}
\item K.I would like to thank C. Gordon for helpful comments. K. T. and K. I. are supported by Grant-in-Aid for the xsJapan Society
 for the Promotion of Science Fellows. N.S. is supported by a
 Grant-in-Aid for Scientific Research from the Japanese Ministry of
 Education (no. 17540276). This work was supported in part
 by the Kavli Institute for Cosmological Physics through the grant NSF
 PHY-0114422.
\end{scilastnote}

\clearpage

\begin{figure}
\centering
\includegraphics[width=0.7\textwidth]{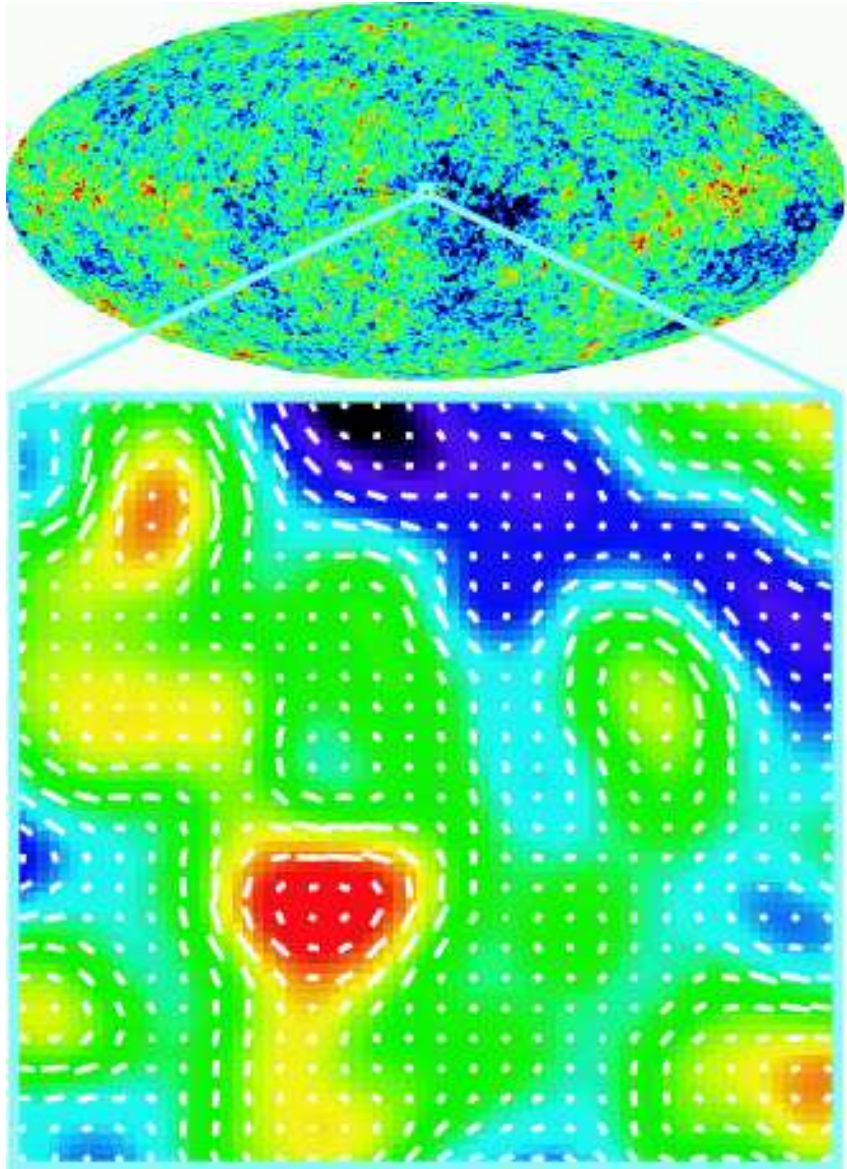}
\caption{All sky map (top) of cosmological microwave background
  anisotropy obtained by the Wilkinson Microwave Anisotropy Probe (WMAP)
  satellite \cite{Bennett:2003bz} and schematic picture (bottom) of
  cosmological magnetic fields generated from density fluctuations
  ($0.5^\circ \times 0.5^\circ$ sky
  field, which corresponds to $130$ Mpc $\times$ $130$ Mpc comoving
  scale). Red regions are hot spots and blue regions are cold spots,
 with a range  
  of temperatures $\sim 2.725$ K $\pm 200 \mu$K.
  The magnetic field vectors are shown together with the map. 
  Strong magnetic fields are generated by currents where
  the gradient of density perturbation in photons is large.}
\end{figure}
\begin{figure}
\centering
\includegraphics[width=0.9\textwidth]{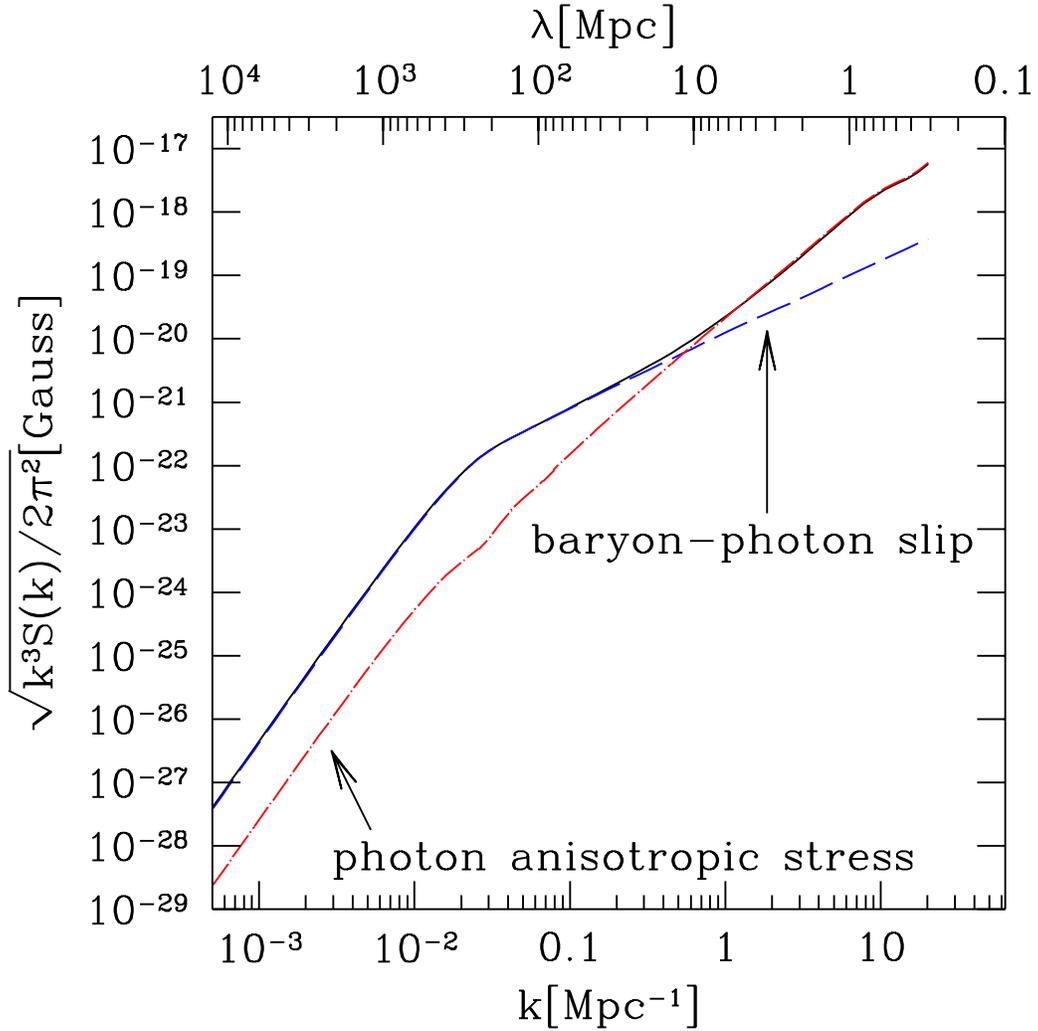}
\caption{Spectrum of magnetic fields S(k) generated from
 cosmological perturbations at cosmological recombination.
 We plotted $\sqrt{[k^3 S(k)]}$ instead of $S(k)$ to measure in units of
 gauss. 
 Blue dashed and red dot-dashed lines show contributions from the
 baryon-photon slip and photon's anisotropic stress (the first 
 and third terms in eq.(S2)), respectively.
 The spectrum decays as $k^4$ at scales 
 larger than that of the cosmic horizon at cosmological recombination. At
 small scales, the contribution from the anisotropic stress of photons
 dominates and the spectrum has a slope proportional to $k$.}
\end{figure}
\begin{figure}
\centering
\includegraphics[width=1.0\textwidth]{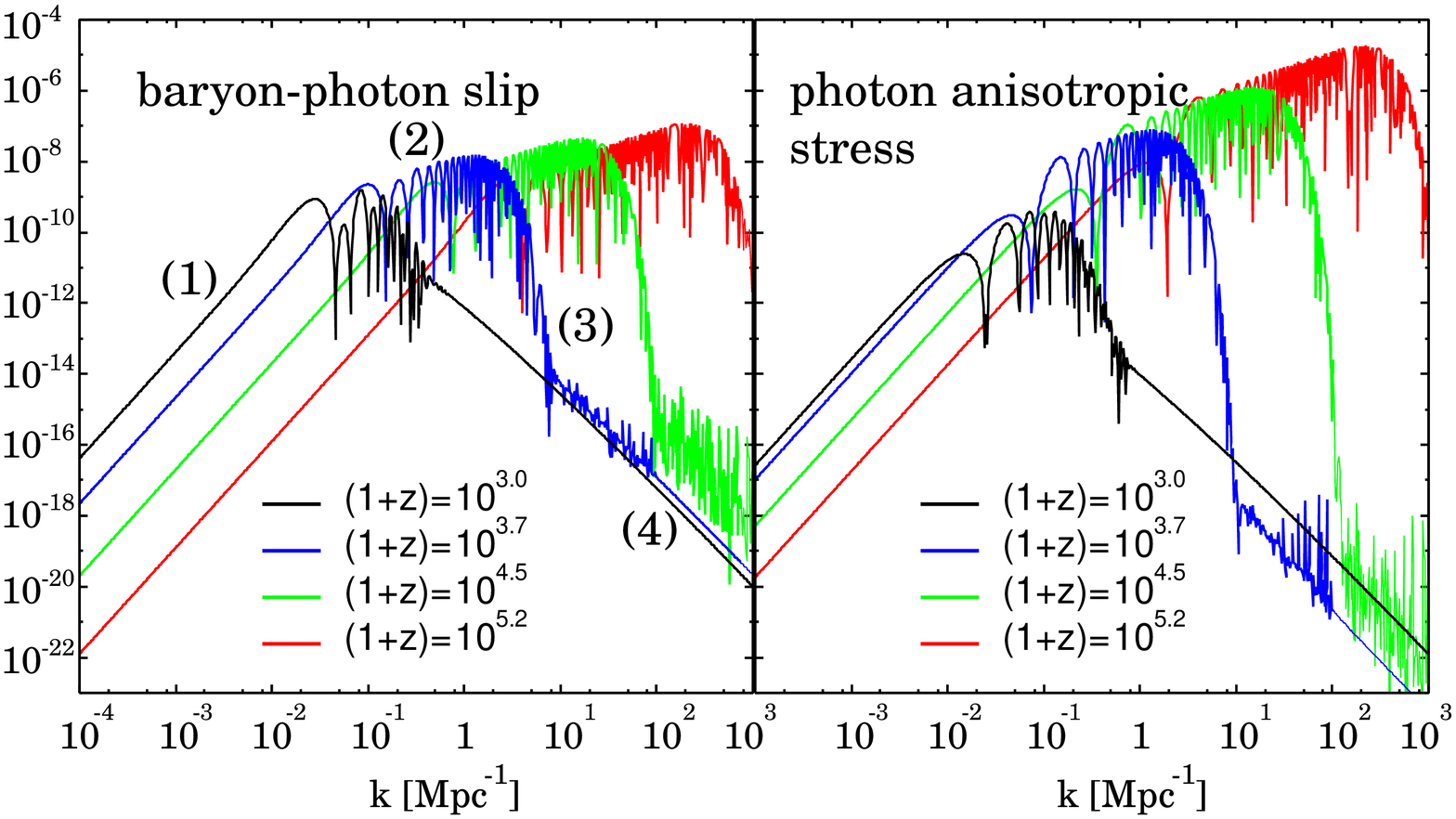}
\caption{Plots of source terms in Fourier space per
 d$\log(1+z)$,  $k^3\frac{\rho_\gamma}{H}\delta_\gamma (v_e-v_\gamma)$
 (baryon-photon slip contribution; left) and
 $k^3\frac{\rho_\gamma}{H}\Pi_\gamma v_e$ (anisotropic stress
 contribution; right), at different redshifts. 
Here, $\rho_\gamma$ is the energy density of photons, $H$ is the Hubble
 parameter, $\delta_\gamma$ is the density fluctuation of photons, $v_e$
 is the bulk velocity of electrons, $v_\gamma$ is the bulk velocity of
 photons, and $\Pi_\gamma$ is the anisotropic stress of photons.
The magnetic field
 spectrum (Fig.2) is obtained by time and 
 $k$-space convolution integrals of these spectra (eq.(S19)).
 The spectrum at each redshift is divided into four parts
 from large scales to small ones (blue line in the left panel): (1)
 featureless primordial power law spectrum 
 at super horizon scales, (2) acoustic oscillation spectrum at
 sub-horizon scales, (3) damping spectrum at diffusion scales, and (4) power
 law phase after the diffusion damping before recombination
 \cite{1998ApJ...501..442Y}. These spectra indicate that magnetic fields
 are created when the modes of perturbations come across the cosmic
 horizon and undergo acoustic oscillations. The redshifts in the figure
 are chosen only for illustration.}
\end{figure}

\end{document}